# Robust estimators for turbulence properties assessment

Daniel Valero[1], Hubert Chanson[2], Daniel B. Bung[3]

[1]Water Science and Engineering, IHE Delft Institute for Water Education, 2611 AX Delft, the Netherlands. E-mail: d.valero@un-ihe.org *(author for correspondence)*
ORCID: 0000-0002-7127-7547

[2]School of Civil Engineering, The University of Queensland, Brisbane, Australia. Email: h.chanson@uq.edu.au

ORCID: 0000-0002-2016-9650

[3]Hydraulic Engineering Section (HES), Aachen University of Applied Sciences (FH Aachen), Aachen, Germany. Email: bung@fh-aachen.de
ORCID: 0000-0001-8057-1193

**ABSTRACT**

Robust estimators and different filtering techniques are proposed and their impact on the determination of a wide range of turbulence quantities is analysed. High-frequency water level measurements in a stepped spillway are used as a case study. The studied variables contemplated: the expected free surface level, the expected fluctuation intensity, the depth skewness, the autocorrelation timescales, the vertical velocity fluctuation intensity, the perturbations celerity and the one-dimensional free surface turbulence spectrum. When compared to classic techniques, the robust estimators allowed a more accurate prediction of turbulence quantities notwithstanding the filtering technique used.

## 1 Introduction

Numerous experimental studies have been conducted in the past decades aiming to determine different turbulence properties of the free surface. With increasing interest on the water surface dynamics, the focus should be placed on its accurate experimental determination. Different methods can be used to describe the free surface, which can be grouped into:

1. Use of classic estimators (e.g., mean, standard deviation, and others).
2. Use of robust estimators (e.g., median, median absolute deviation, and others).

Traditionally, classic estimators have been the preferred choice. These statistical methods are best for stringent situations where raw data perfectly match physical characteristics. Nonetheless, true data are commonly contaminated by outliers and noise inherent to any experimental methodology, and filtering should be prescribed to reduce the impact on the turbulence estimations. Alternatively, robust estimators can be used, understanding "robust" or resistant as the characteristic to be affected only to a limited extent by a number of gross errors (Hoaglin et al. 1983). This is achieved when a small subset of the sample cannot have a disproportionate effect on the estimate. In turn, the robust estimators may not be distribution-free estimators, although they are best for a broad range of situations, tolerate a large quantity of outliers mixed within the data sample (up the breakdown point) without resulting in a meaningless estimate, and perform superiorly even for small data samples (Hoaglin et al. 1983).

A simple example can be presented through the estimation of the expected value of a variable. As a dataset contains outliers, the use of the mean (classic) estimators should be preceded by a filtering step. Otherwise, a median (robust) estimator can be used, being more insensitive to the presence of outliers. When outliers are *obvious*, they can still be removed but accurate determination of the expected value does not strongly rely on the adequacy of the filtering technique.

This work explores the aforementioned dual data analysis by studying different combinations of filtering techniques and classic/robust estimators for an extremely turbulent flow case: the turbulent free surface in the non-aerated region over a stepped spillway (introduced in Section 2). Robust estimators, mainly based on the simple concept of median and data ranking, are proposed in Section 3 for a wide range of turbulence properties, namely: the expected free surface level, the expected fluctuation intensity, the depth skewness, the autocorrelation timescales, the vertical velocity fluctuation intensity, the perturbations celerity and the one-dimensional free surface turbulence spectrum. Alternatively, three filtering techniques based upon well-stablished works (Goring and Nikora 2002, Wahl 2003) are presented in Section 4. These three techniques present gradually increasing intricacy, encompassed of higher rejection rates. Section 5 analyses the combination of classic and robust estimators with the proposed filtering techniques for different turbulence quantities of the turbulent free surface over a stepped spillway. Results discussion and final conclusions are presented in Sections 6 and 7, respectively.

## 2 Experimental setup

### 2.1 Geometry

The present study focuses on the non-aerated region of a large stepped spillway model of 45º slope (1V:1H) located at the University of Queensland. The stepped spillway has a wide inlet basin (5 m wide, 2 m long) which ensures smooth inlet conditions, leading to a broad crested weir (0.60 m long, 0.985 m wide) which conveys the flow into the stepped spillway (same width) composed of steps of height $h = 0.10$ m. The water discharge is estimated using a previously obtained an experimental discharge relationship based on detailed velocity measurements. A thorough description of this case study, and main flow variables, can be found in Zhang and Chanson (2016). The experimental setup is shown in Fig. 1.

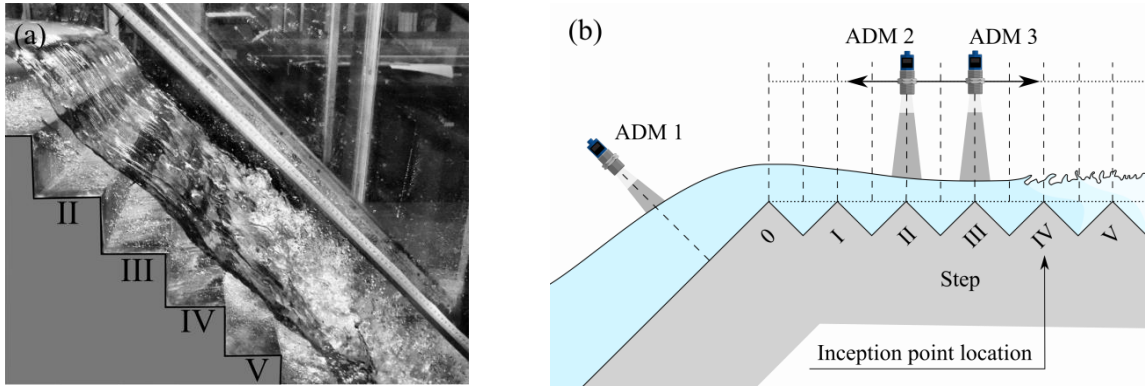

Figure 1.Stepped spillway. (a) Image taken at 1/1000 shutter speed, processed with contrast-limited adaptive histogram equalization to enhance the characteristic free surface perturbations and air entrainment after step IV. (b) Sketch of the experimental setup (rotated 45° counterclockwise), steps numbering, sensors measurements location (- -), pseudobottom and parallel axis (:). Flow from left to right.

*2.2   Instrumentation*

Instantaneous free surface measurements were sampled with three microsonic™ Acoustic Displacement Meter (ADM) mic+25/IU/TC. The measuring range recommended by the manufacturer is 30 – 250 mm. The near field of the ADM sensors was enclosed with PVC cylinders of the same diameter to prevent the wetting of the sensors. This artefact did not alter the sensors' output signal, as shown by Kramer and Chanson (2018).

The ADM sensors provide a voltage time series which can be correlated to a distance in order to estimate a water level. The three ADM sensors were calibrated over a distance range covering the expected water depths to be measured. Calibration was conducted by recording during 300 s at 100 Hz for 11 different distance levels, which covered the range of expected flow depths. Figure 2a shows that calibration exhibited a linear relation over the entire sampled range (note that different locations of the sensors yield different voltage-depth relationships) and Fig. 2b that the Standard Deviation (STD) of each calibration step remained close to 0.10 mm, computed from the voltage STD and using the obtained calibration curve, which matches the accuracy specified by the sensors manufacturer.

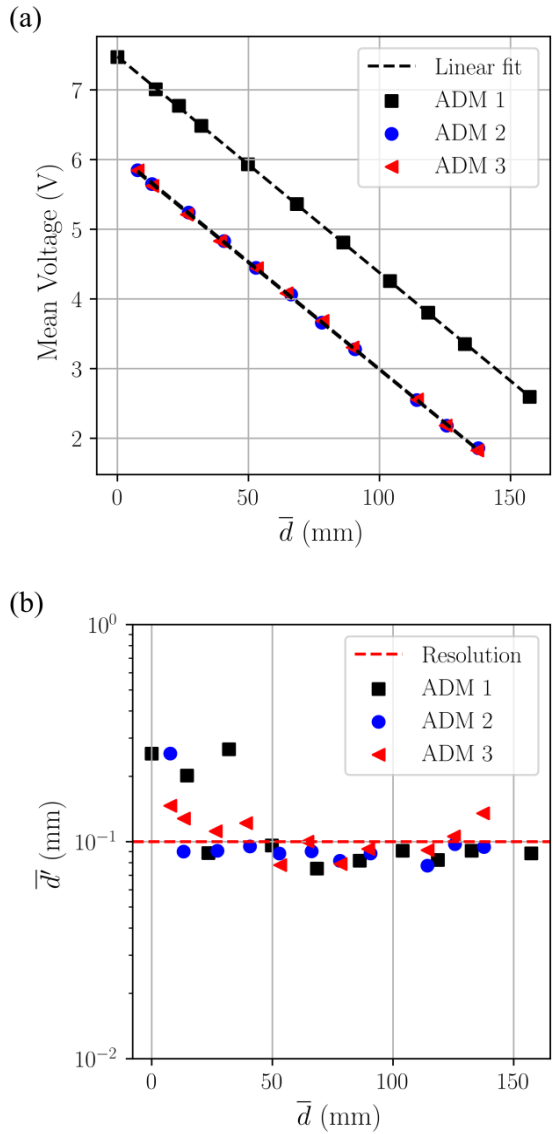

Figure 2. ADM performances. (a) ADM calibration curves (mean flow depth $\bar{d}$ and corresponding mean voltage level) and (b) Depth fluctuation $\bar{d'}$ for static measurements (noise level) corresponding to each mean flow depth level measured during the ADM calibration curve determination. 'Resolution' specified by sensors' manufacturer.

*2.3    Measurement location and flow conditions*

The ADM 1 was located at a fixed position over the crest, 0.17 m upstream over the edge of the first step (step 0, Fig 1b). The other two sensors (ADM 2 and ADM 3) were located over the stepped geometry separated by 0.141 m in the longitudinal direction, thus coinciding with one cavity length. Keeping a constant distance between both sensors, ADM 2 and ADM 3 were placed above the pseudobottom (formed by the step edges, Fig. 1b), allowing the measurement of the flow depths at different spillway locations. Recordings were conducted at the step edges (steps 0 – VII) and above the step cavities (mid distance between the step edges), as marked in Fig. 1. Each recording was conducted at a sampling rate of 100 Hz during 600 s. The total time recorded and the distribution of the measurements over the spillways is shown in Fig. 3. Differences of sampling time are due to two reasons: overlapping of measurement locations as the ADM sensors were moved downstream; and repetition of some measurements at locations where the free surface was highly roughened due to turbulence. The inception point of air entrainment is marked for each investigated discharge ($d_c/h$ = 0.9, 1.1, 1.3, 1.5, 1.7, 1.9 and 2.1, with $d_c$ the critical depth) according to the visual observations of Zhang and Chanson (2016). For reference, the empirical formulas of Meireles et al. (2012) and Chanson et al. (2015) are included. Precisely, all the measurements fall within the non-aerated region, where the flow gradually roughens as the flow becomes more turbulent up to break up (Valero and Bung 2018).

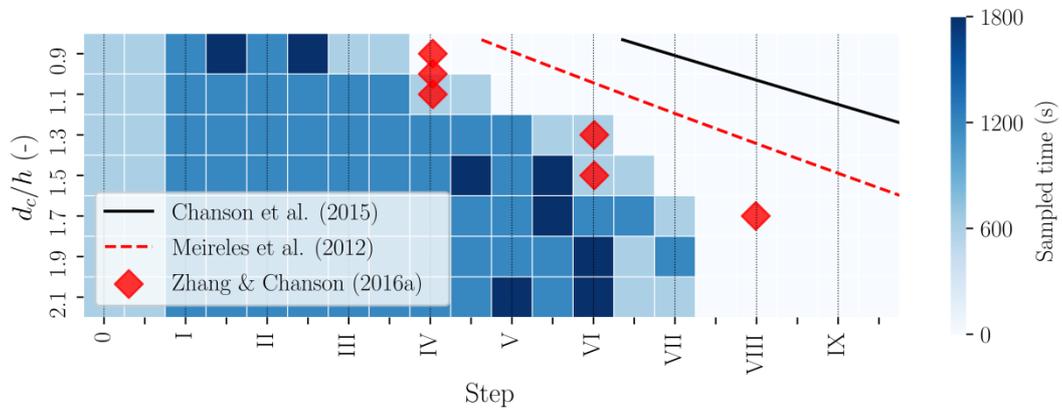

Figure 3. Measurements distribution and sampled time for the studied flow conditions. Visual observation of the inception point location (Zhang and Chanson, 2016) and empirical formulas of Meireles et al. (2012) and Chanson et al. (2015).

## 3 Robust estimators

The use of estimators insensitive to the presence of outliers can yield more reliable turbulence predictions, hence alleviating the responsibility often relying solely on the filtering techniques. In the following, the estimators corresponding to classic statistic techniques are presented overlined (e.g., $\bar{d}$) whereas the robust counterpart is presented with a tilde (e.g., $\tilde{d}$).

### 3.1 Expected value and fluctuation

The expected value of a variable (E[·]) is the first variable of interest in any data analysis. For the case of the flow depths, it is often estimated by using the mean ($\bar{d}$), defined as the ensemble average of all the samples in the filtered signal:

$$\mathrm{E}[d] := \bar{d} = \frac{1}{N}\sum_{i=1}^{N} d_i \qquad (1)$$

with $N$ the total number of flow depth measurements. Deviations from the expected value of a variable are also of interest, as they are associated to turbulence and dispersion, and can be studied on the basis of:

$$\eta = d - \mathrm{E}[d] \qquad (2)$$

By definition, $\mathrm{E}[\eta] = 0$ and it is frequent to study the expected value of the squared fluctuation. An estimation of the dispersion of the data can be done by means of the sample standard deviation (STD) as:

$$\sqrt{\mathrm{E}[\eta^2]} := \bar{d}' = \sqrt{\frac{\sum_{i=1}^{N}(d_i - \bar{d})^2}{N-1}} \qquad (3)$$

The STD approximates the population variance using the squared value of each sample deviation, thus endorsing bigger weight to the outliers which depart significantly from the expected value of the series. The mean value can be affected by the presence of outliers as well but, when equally distributed around the mean, their contribution to Eq. (1) would balance. Alternatively, the median (MED) and the Median Absolute Deviation (MAD) can be used as estimators of location and variance, being both robust estimators against outliers with a breakdown point of 50 % (i.e., 50 % of contaminated data is necessary to force the estimator to result in a false output) as opposed to the counterpart mean and standard deviation, which hold a 0 % breakdown point. It is noteworthy that the median is the location estimator that presents the highest breakdown point (Leys et al. 2013) and is defined as the value separating the greater and lesser halves of the series. Hence, a robust estimator for the expected value is herein proposed directly through the median operator $\tilde{d} = \mathrm{MED}(d)$. Similarly, the MAD represents the best robust scale estimator, even more than the interquartile range that remains at a 25 % breakdown point (Rousseeuw and Croux 1993; Leys et al. 2013).

The MAD can be obtained by sorting the absolute value of the residuals around the MED and selecting the value corresponding to the 50 %. Nonetheless, it is implemented in many commonly used numerical libraries (e.g., MATLAB®, R programming language or Python 2.7 together with the *statsmodels* library, being the latter combination the one used in this study). The MAD of the sampled flow depth can be related to the standard deviation of different probability density functions as (Rousseeuw and Croux 1993):

$$\tilde{d}' = k\,\mathrm{MAD}(d) = k\,\mathrm{MED}(|\eta|) \qquad (4)$$

being $k$ a coefficient related to the sample distribution. When a Gaussian behaviour is assumed, $k$ takes the value (Rousseeuw and Croux 1993):

$$k = 1.483 \tag{5}$$

Another estimation of the flow depth variance can be obtained through the interquartile range $\widetilde{d_I}' = \widetilde{d_{75}} - \widetilde{d_{25}}$, being $\widetilde{d_{75}}$ and $\widetilde{d_{25}}$ the depth levels representing the 75$^{th}$ and 25$^{th}$ quartiles, respectively. This estimator presents a lower breakdown point than the MAD, therefore being more sensitive to the presence of outliers.

*3.2  Skewness*

Higher order statistics can be computed to study the shape of the free surface waves. With increasing order, the exponent weighting the outliers is also incremented although for even order numbers some positively and negatively deviated outliers could balance.

The flow depth skewness ($\overline{d''}$) can be defined as:

$$\overline{d''} = \frac{1}{N}\sum_{i=1}^{N}\left(\frac{d_i - \bar{d}}{\bar{d}'}\right)^3 \tag{6}$$

This descriptive statistic is a dimensionless measure of the lack of symmetry. Following the robust estimators defined for first and second order statistics, quartiles information can be used to define a robust estimator for the skewness (Zwillinger and Kokoska 2000):

$$\widetilde{d''} = \frac{\widetilde{d_{75}} - 2\,\breve{d} + \widetilde{d_{25}}}{\widetilde{d_I}'} \tag{7}$$

The parameter $\widetilde{d''}$ is the so-called quartile coefficient of skewness (Zwillinger and Kokoska 2000).

*3.3  Autocorrelation timescales*

The cross-correlation function between two series ($x$ and $z$) can be computed as:

$$R_{xz}(\tau) = \frac{\text{cov}(x, z(\tau))}{\sqrt{\text{E}[(x - \text{E}[x])^2]}\sqrt{\text{E}[(z(\tau) - \text{E}[z])^2]}} \tag{8}$$

being cov the covariance and $\tau$ the lag time. Efficient computation of the covariance for long samples can be achieved through fast Fourier transformation. Note that some terms of Eq. (8) can be rewritten using classic estimators as:

$$\overline{R_{xz}}(\tau) = \frac{\text{cov}(x, z(\tau))}{\overline{x'}\ \overline{z'(\tau)}} \tag{9}$$

An alternative, nonparametric form of the correlation can be defined by means of the Spearman's correlation. For this purpose, both $x$ and $z$ are ranked separately from smallest to largest values, assigning the mean rank when equal values occur. Let $u_i$ and $w_i$ take the rank of the $i^{\text{th}}$ observation in $x$ and z, respectively. Spearman's rank correlation of $x$ and $z$ can then be computed as (Zwillinger and Kokoska 2000):

$$\widetilde{R_{xz}}(\tau) = \overline{R_{uw}}(\tau) = \frac{\text{cov}(u, w(\tau))}{\overline{u'}\ \overline{w'(\tau)}} \tag{10}$$

The use of the ranked data has the advantage that it allows computation of the correlation between both trends without strongly depending upon the current value of each measurement. This alternative correlation does not considerably slow down the computation as the ranking of the vectors can be done with efficient sorting algorithms and, afterwards, the sorted data can be correlated as per the original vectors. Spearman's correlation is the nonparametric version of the Pearson correlation coefficient, which makes its computation more robust to outliers.

Cross-correlation function can be used to obtain the most probable lag between two time series [= argmax($R_{xz}$)] or to obtain the autocorrelation function ($R_{xx}$), that allows extraction of turbulent scales and spectrum. A timescale ($T_{\eta\eta}$) for the flow depth fluctuations can be obtained by integrating its autocorrelation function ($R_{\eta\eta}$) up to the first zero-crossing point:

$$T_{\eta\eta} = \int_0^{\tau=\tau(R_{\eta\eta}=0)} R_{\eta\eta}(\tau)\ \mathrm{d}\tau \tag{11}$$

Depending on how the autocorrelation function is computed (Eq. 9 or 10), a classic estimation of the turbulent timescale ($\overline{T_{\eta\eta}}$) could be carried out through $\overline{R_{\eta\eta}}$ or, an alternative, robust turbulent timescale ($\widetilde{T_{\eta\eta}}$) could be computed through $\widetilde{R_{\eta\eta}}$.

*3.4 Cross-correlation peak and wave celerity*

Given that two ADMs are placed in series (ADM 2 and ADM 3), their signals can be cross-correlated allowing estimation of the most probable time lag. When the distance between the synchronized sensors is known, the celerity of the free surface perturbations can be estimated. Both the classic correlation (Eq. 9) and the Spearman's correlation (Eq. 10) can be used to compute the cross-correlation and, hence, estimate the waves' celerity ($\bar{c}$ and $\tilde{c}$, respectively).

*3.5  One-dimensional flow depth spectrum*

The one-dimensional spectrum has been traditionally computed for velocity time series. It allows insight into the flow structure and the energy distribution for different wavelengths (i.e., different eddy sizes). In this study, the one-dimensional spectrum, as defined by Pope (2000), is proposed for the flow depth fluctuation:

$$E_{\eta\eta}(f) = \frac{2}{\pi} \, E[\eta^2] \int_0^\infty R_{\eta\eta}(s) \, \cos(f\,s) \, \mathrm{d}s \qquad (12)$$

Both $E[\eta^2]$ and $R_{\eta\eta}$ can be estimated either in a classic (Eqs. 3 and 9) or a robust manner (Eqs. 4 and 10), thus leading to a standard ($\overline{E_{\eta\eta}}$) or robust ($\widetilde{E_{\eta\eta}}$) estimation of the one-dimensional depth fluctuation spectrum.

# 4  Data filtering

A data point is oftentimes labelled as an outlier when it lies at an abnormal distance from other values of a certain population. Nonetheless, for any observation far from the group, there is a positive (despite small) possibility to occur and thus the crux is on identifying these outliers without losing true information from the population (Barnett and Lewis 1978). Doubtful or anomalous values can come from a mixed sample of a different population or erroneous measurements. It is also important to understand how outliers are physically generated to understand their likelihood of occurrence.

*4.1  Lower and Upper Voltage (LUV)*

When an ADM pulse-echo is lost, the sensor is incapable of generating a proper estimation of the free surface position. The voltage provided for these lost echoes usually piles at the highest or lowest voltages, far away from the voltage values corresponding to realistic depths. It is therefore convenient to locate the sensor so that the measurements are contained in a region of interest far from the extreme voltage values. A first filtering approach could be to simply remove values below and above 5 % and 95 % from the total voltage range of the ADM. This double threshold filtering technique is based on a physical observation. For the ADM model used in this study, the voltage filtering levels correspond to 0.5 V and 9.50 V respectively (Fig. 4a).

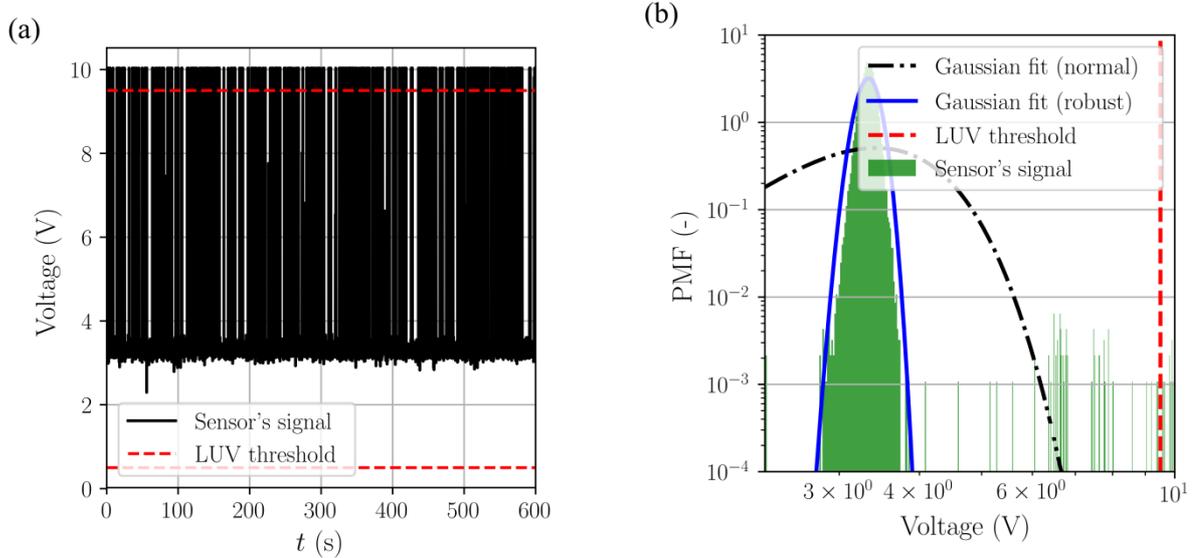

Figure 4. Signal of ADM 2, $d_c/h = 2.1$ and step V. The number of outliers removed using LUV technique corresponds to 4.8 % of the total number of samples. (a) Sensor's signal (temporal series); (b) voltage levels distribution and Gaussian fit using classic (normal) and robust estimators.

*4.2   Robust Outlier Cutoff (ROC)*

Outliers depart from the expected estimation of the flow depth, but do not necessarily accumulate out of the LUV bounds. The Probability Mass Function (PMF) shows that some erroneous measurements run together at different voltage levels. A quick flow observation indicates that these voltage values associated to different water levels are not physically meaningful and the introduction of narrower bounds arises as a preferred alternative than LUV filtering.

A commonly used technique is to estimate the variance of the sample, by means of the STD, to establish the filtering bounds around a certain number of STD away from the mean. An alternative way to estimate variance can be done through robust estimators, as presented in Eq. 4. Difference between normal and robust estimators can be well-perceived in Fig. 4b, where a Gaussian function is fitted using the location and variance obtained with the classic estimators (mean and STD) and through the robust estimators (MED and MAD). Figure 4b also shows a small proportion of outliers piling up at different voltage levels, which are readily observable when using a vertical log-scale.

On the question of how many standard deviations are necessary to be accounted for to make sure that "good data" is not filtered out, the universal threshold represents a conservative estimator. It can be expressed as (Goring and Nikora 2002):

$$\lambda_u = \sqrt{2 \ln(N)} \qquad (13)$$

with $N$ the total number of data points of the sample. Use of the universal threshold yields bounds wide enough to avoid filtering out good data, even if the underlying distribution is slightly skewed, but (usually) narrower bounds than those proposed by the LUV technique. If the final distribution is markedly skewed, MAD can be estimated for both positive and negative deviations departing from the MED value and, consequently, different filtering thresholds could be defined for positive and negative deviations.

*4.3    Depth-Velocity Elliptical Despiking (DVED)*

One step further on the filtering of the flow depth time series could be conducted using the finite differences of $\eta$ and its variance, following Goring and Nikora (2002) for velocity data. Provided that an ADM can measure the time series of flow depth, a vertical velocity ($v_s$) can be estimated by using the central finite difference:

$$v_s \equiv \frac{\partial d}{\partial t} = \frac{\partial \eta}{\partial t} \approx \frac{\Delta \eta_j}{\Delta t_j} = \frac{\eta_{j+1} - \eta_{j-1}}{t_{j+1} - t_{j-1}} \quad (14)$$

For the data at the beginning and ending of the sample, backward or forward differences can be taken by simply using $j$ instead of $j+1$ or $j-1$ at the right-hand side term of Eq. (14).

For $\eta$, maximum and minimum thresholds can be established based on Eqs. (3) and (14) for $v_s$ as well. This approach is a simplified version of the work of Goring and Nikora (2002) and Wahl (2003), which extended the analysis up to the second derivative of the variable under analysis. It must be noted that the filtered data could still produce unrealistic vertical velocities, because some depth outliers could randomly fall inside the bounds defined by the ROC technique, hence remaining undetected.

Goring and Nikora (2002) proposed ellipsoid-type bounds based on the observation that "good data" tend to cluster together forming this shape. In this study, the 2D PMF for $\eta$ and $v_s$ of the data was analysed and similar clustering forms were recognized. For conciseness, only the 2D PMF of the data previously shown in Fig. 4a and 4b is presented in Fig. 5. The isoprobability contours (points with same probability of occurrence) appear to take the form of ellipse-like curves. Accounting for the universal threshold as a situation of equal likelihood of appearance, the hypothesis of Goring and Nikora (2002) remains consistent for flow depths measurements. Thus, the filtering bounds can be finally expressed as:

$$\left(\frac{\eta}{MAD(\eta)}\right)^2 + \left(\frac{v_s}{MAD(v_s)}\right)^2 \leq (\lambda_u k)^2 \quad (15)$$

Equation (15) reduces to the ROC filtering when the second term of the left-hand side is neglected.

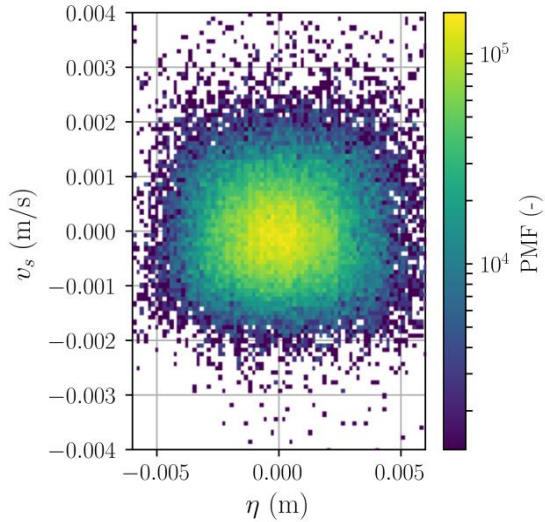

Figure 5. Probability Mass Function (PMF) of the data shown in Fig. 4.

*4.4 Practical implementation considerations*

The three proposed filtering methods correspond to: one physically based filtering (LUV), and two statistical techniques (ROC and DVED). The ROC and DVED algorithms produce upper and lower bounds based on the quartiles of the sampled data, while the LUV method defines bounds based on the physical observation that extremely high and low voltages correspond to lost echoes. Thus, LUV can be complementary to ROC and DVED techniques under certain conditions. An example can be given by a recording for which the free surface is considerably sloped relative to the axis of the ADM resulting on more than 50 % of erroneous data. In such case, most measurements would fall close to 10 V and the application of ROC or DVED alone would not result in adequate identification of the outliers. Hence, it is proposed that, after applying ROC and DVED, the LUV technique should be used to avoid accepting erroneous data. In the following, any result presented for ROC and DVED techniques has been also filtered using LUV afterwards. Additionally, if a filtering technique flagged more than 50 % as invalid data, the recording was dismissed and not accounted in the subsequent analysis.

When an outlier is detected by any of the presented filtering methods, it can be simply deleted or replaced. The most basic method would consider the removal of the outlier. The outlier could also be substituted by the average or the median of the entire signal, which would imply that the value replacing the outlier simply takes the "expected" value of the signal. Nonetheless, this can create fast gradients which were not contained in the original signal and, it could considerably affect the latter computed statistical estimators (e.g., by reducing the STD). In this study, it is proposed to substitute the outliers by linear interpolation between their surrounding points. More complex methodologies could be proposed based on involving a larger number of points or accounting for the autocorrelation of the signal to generate the outlier replacement.

## 5 Results

*5.1 Rejection rate*

The three proposed filtering techniques were applied to the data presented in Fig. 3, obtaining different rates of rejection for different measuring locations (see Fig. 6 and Table 1). Figure 6 shows that large percentages of data were rejected close to the spillway crest (step 0), where free surface bends following the chute axis at the transition from the broad crested weir to the spillway (Fig. 1). This can be explained by the inclination of the detection zone axis of the ADM with respect to the normal to the free surface (Zhang et al. 2018). These large rejection rates close to the first step were obtained for all discharges, indistinctly of the filtering technique.

With increasing discharge, the flow depth becomes more parallel to the pseudobottom, as opposed to the flow depth curvatures that encompass the step edges that are observed for the lower discharges. Hence, the free surface tends to be closer to the axis of the ADM measuring cone at large discharges and fewer outliers can be expected. Close to the inception point of air entrainment, the free surface considerably roughens (Valero and Bung 2018) and its dynamic determination can be more challenging for the ADM sensors, consequently resulting in a local increase of the outliers contained in the recorded dataset.

Generally, no major change in the amount of rejected data occurred when applying the ROC technique compared to the LUV method, but approximately twice more data was removed when applying the DVED technique (see Table 1).

Table 1. Median percentage of rejected data through the non-aerated region of the spillway.

|     | Filtering method | | |
| --- | --- | --- | --- |
| ADM | LUV (%) | ROC (%) | DVED (%) |
| 2 | 3.0 | 3.2 | 8.3 |
| 3 | 5.7 | 6.2 | 12.2 |

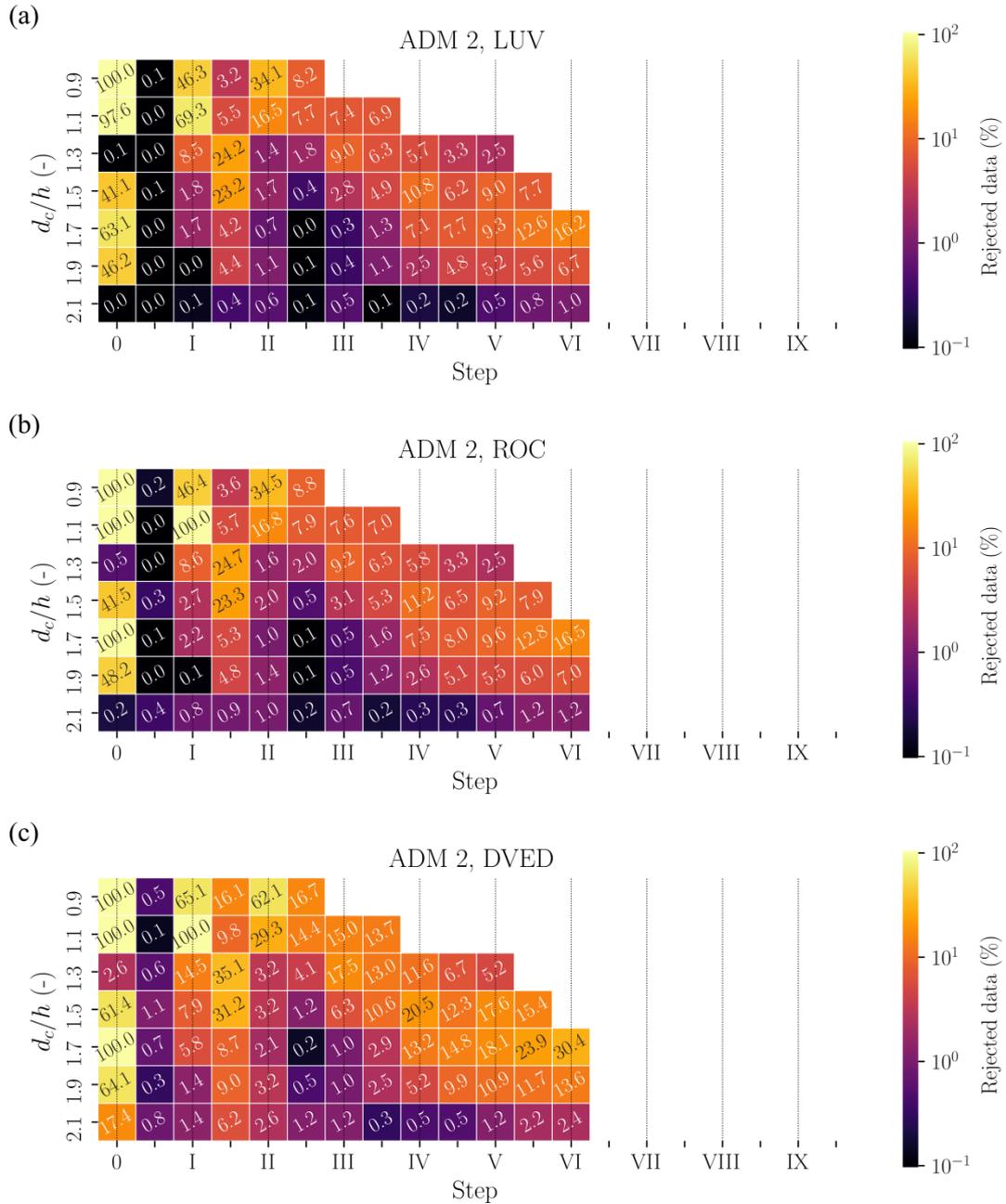

Figure 6. Percentage of rejected data from ADM 2 after application of different filtering techniques: (a) LUV, (b) ROC and (c) DVED.

## 5.2 *Expected value of the flow depth and its fluctuation*

The first analysed variable corresponds to the expected value of the flow depth. A similar prediction with the mean estimator (Eq. 1) is obtained independently of the filtering method used. Positive and negative deviations compensate, resulting in a negligible effect on the mean estimation for almost all the data. For the median estimator, all data points yield an exact same estimation, independently of the filtering technique used. Consequently, both mean and median estimators can provide with accurate values of the expected flow depth regardless of the filtering technique.

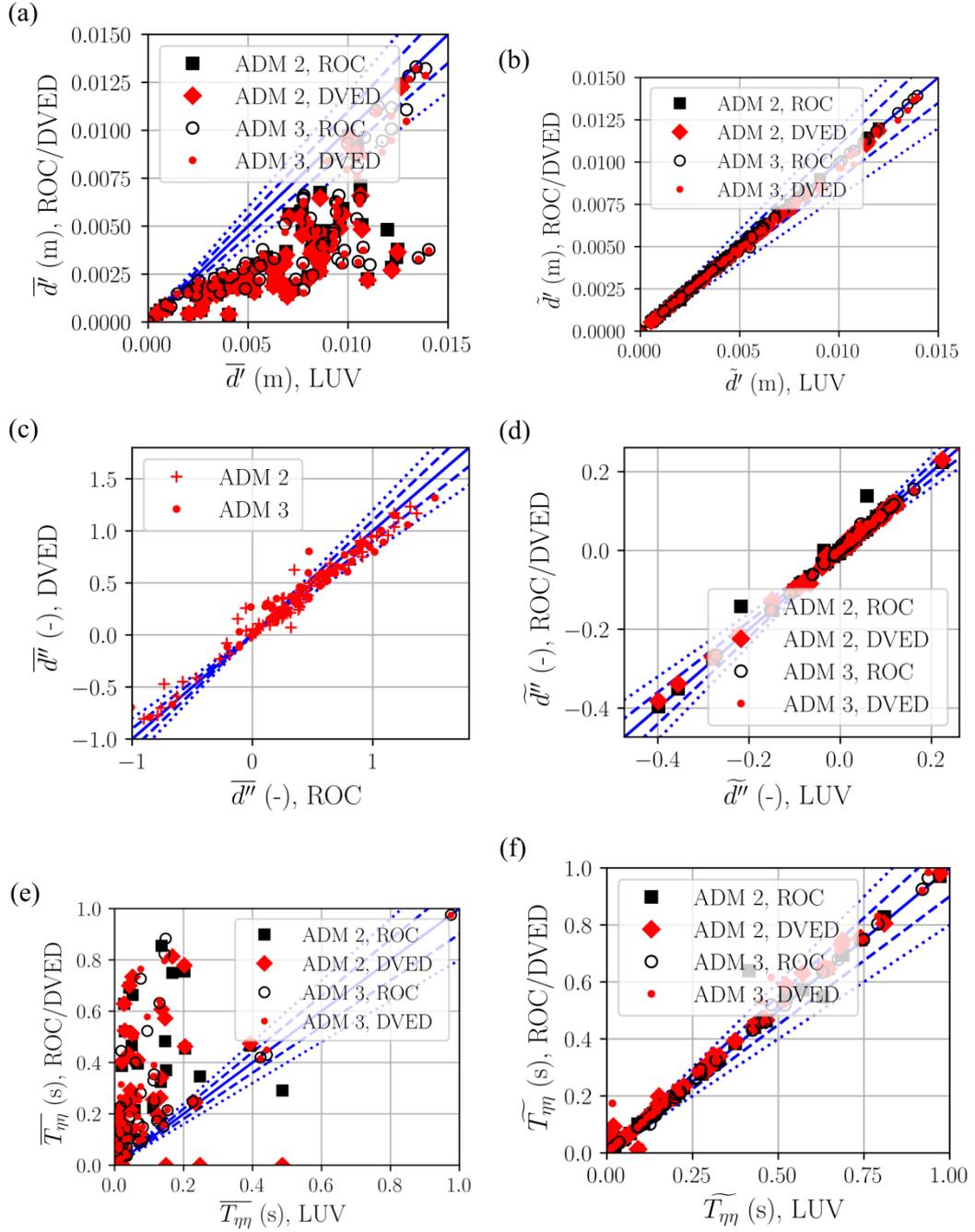

Figure 7. Effect of different filtering techniques on different turbulent flow variables using traditional (left) and robust (right) estimators. (a, b) expected value of the depth fluctuation, (c, d) skewness and (e, f) turbulent timescales. Note that for (a-f) ROC and DVED perform likewise and that (c) does not include LUV filtered data (which showed random, out of order spreading). Perfect agreement (—), +/- 10 % deviation (- -) and +/- 20 % deviation (⋯).

In terms of the mean fluctuation of the flow depth (Eq. 3), a significant deviation from the prediction of $\overline{d'}$ is obtained by using the filtered data series using the LUV method as well as the method using the filtered data series of the ROC and the DVED methods (Fig. 7a). In Figure 7a, both the ROC and the DVED filtered data converged to similar values, with differences below 10 %. It is then proposed that when studying the depth variance, at least the ROC method should be used, instead of the LUV method, to avoid incorporating outliers as an unphysically higher turbulence level. An alternative approach, if the analysis is just restricted to mean and mean fluctuation levels, is to filter the data using the LUV method but taking advantage from the robust behaviour of the MAD estimator (Eq. 4) to approximate the samples' expected fluctuation. In such case, comparison of $\widetilde{d}'$ estimation between the LUV and the DVED filtered data (Fig. 7b) revealed that roughly all the data falls between perfect agreement and $-10\ \%$ lines, as the it occurs for the ROC and DVED filtered data for $\overline{d}'$ (Fig. 7a). Performance for $\widetilde{d_I}'$ was in close agreement to that of $\widetilde{d}'$ (shown in Fig. 7b), with almost all the predictions varying less than 10 % indistinctly of the filtering method.

### 5.3 Flow depth skewness

Discrepancies between $\overline{d''}$ estimated based upon data obtained after filtering with the ROC and the DVED algorithms and the classic skewness estimator (Eq. 6) remained usually below 20 %. Using only the LUV filtering method, the remaining outliers scaled up by two orders of magnitude the skewness predictions. This is shown in Fig. 7c, as a comparison between the results obtained after filtering with the ROC and the DVED algorithms and the classic skewness estimator (Eq. 6).

Figure 7d shows the comparison for all three filtering techniques using robust estimators (Eq. 7). It must be noted that $\widetilde{d''}$ is restricted to values between -1 and +1, which prevents from direct comparison to $\overline{d''}$. Nonetheless, a similar trend can be observed for both classic and robust estimators, with the latest allowing reasonable skewness estimations even for the less restrictive filtering technique (LUV).

### 5.4 Autocorrelation timescales

Autocorrelation timescales estimated with the classic estimators (Eqs. 9 and 11) are considerably smaller when the LUV method is applied, instead of the ROC or DVED (Fig. 7e). Likewise, data filtered with the ROC technique led to predictions 20 % smaller for $\overline{T_{\eta\eta}}$ than the when using the DVED method (Fig. 7e). Results for the robust estimation of the autocorrelation timescale ($\widetilde{T_{\eta\eta}}$, Eqs. 10 and 11) are shown in Fig. 7f. Using the robust Spearman's ranked autocorrelation, the differences in the estimation of the autocorrelation timescales reduce significantly between the data filtered using the LUV method and the other more stringent filtering techniques. Figures 7e and 7f show that the effect of noise in the sampled data is to reduce the autocorrelation function values, which yields estimations corresponding to shorter or faster eddies. For all cases, the numerical integration of Eq. 11 was conducted using the trapezoidal rule.

### 5.5 Vertical velocity fluctuation

An estimation of the instantaneous free surface vertical velocity can be obtained using Eq. (14), with a zero-mean value in a steady flow. The intensity of the vertical velocity fluctuation can be studied in terms of the STD of the velocity time series ($\overline{v_s'}$). The resulting $\overline{v_s'}$ values obtained from the filtered data of all three proposed techniques follow closely the same trend as results for $\overline{d}'$. The data filtered using the LUV method produces considerably larger values of $\overline{v_s'}$ than the other two filtering techniques, as some noise is incorporated as a turbulence level. When taking advantage of the robust nature of the MAD to estimate $\widetilde{v_s'}$, similar results for all three filtering techniques are obtained indistinctly of the filtering method; likewise for $\widetilde{d'}$, shown in Fig. 7b. Hence, discrepancies in the robust estimation between the LUV and DVED filtered data remain around + 20 % deviation and the perfect agreement lines.

*5.6 Free surface perturbation celerity*

Figure 8a shows that the data filtered using the two most stringent methods, ROC and DVED methods, yields similar $\bar{c}$ estimations. Differently, estimations based on the LUV filtered data scatter significantly. Maximum correlation ($\overline{R_{max}}$) is also shown in Fig 8a as it is an indicator of similarity between the two cross-correlated signals. The median value for all measurements of $\overline{R_{max}}$ was 0.16 for the LUV filtered data, 0.45 for the ROC filtered data and 0.44 for the DVED filtered data; showing a clear improvement with the most restrictive filtering techniques.

Using the Spearman's based cross-correlation method (Eq. 10), the estimations generally relied between the +/− 20 % accuracy range, even for the LUV and DVED filtered data (Fig. 8b). Both the ROC and the DVED filtered data estimations coincided for most measurements. The maximum correlation $\widetilde{R_{max}}$ was 0.42 for the LUV filtered data, 0.44 for the ROC filtered data and 0.41 for the DVED filtered data; showing similar levels than after applying the ROC filtering technique with the classic correlation approach.

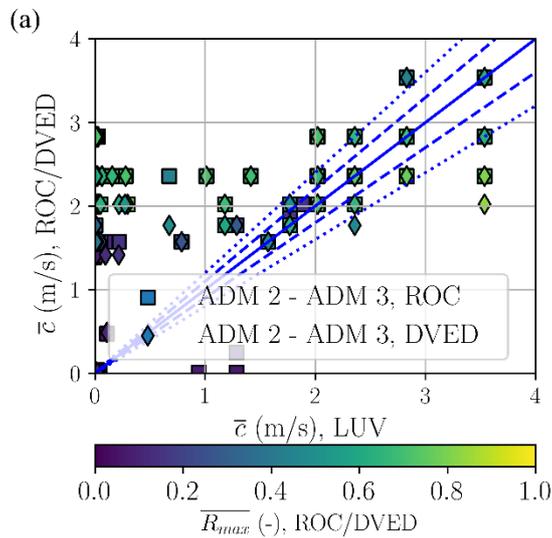

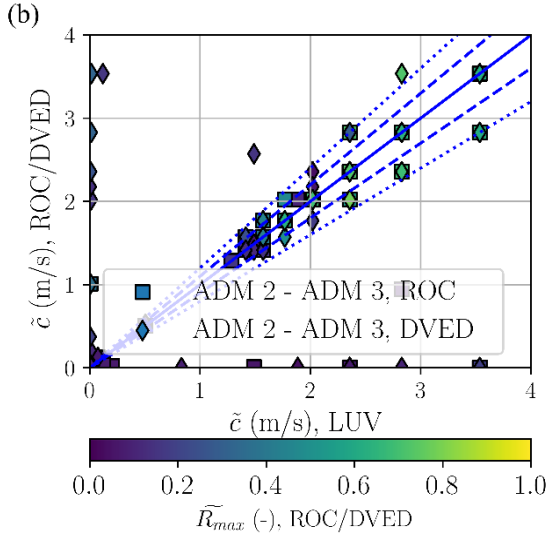

Figure 8. Effect of different filtering techniques on the free surface perturbations celerity obtained with: (a) traditional cross-correlation ($\bar{c}$) and (b) with Spearman's ranked cross-correlation ($\tilde{c}$). Perfect agreement (—), +/- 10 % deviation (- -) and +/- 20 % deviation (···).

*5.7 One-dimensional flow depth fluctuation spectrum*

Spectrum shown in Fig. 9 were obtained by dividing the 600 s samples into 60 equal length non-overlapping signals; which is long enough based on results of Fig. 7e,f for the autocorrelation timescales (which held values generally around 0.5 s). The resulting spectra were subsequently ensemble-averaged. This procedure is based on that proposed by Welch (1967), despite the temporal window has been chosen arbitrarily large as to comprehend a wide range of turbulent timescales. The one-dimensional spectra were obtained for different flow conditions (Fig. 3) but, for the sake of briefness, the effect of the three filtering techniques and the use of classic and robust estimators is only shown for one step and flow condition, albeit it was representative of all others.

Figure 9a shows that all three filtering techniques together with classic estimators yield similar power law slopes, although the application of the LUV technique results in a spectrum with higher energies – associated to considerably higher noise levels – as expected from the results for the flow depth fluctuation estimation. Smoother spectra are obtained when filtering with the most restrictive filtering techniques (ROC and DVED methods). Both the ROC and the DVED filtering data produced a very similar spectrum. Figure 9b shows that all three spectra based upon robust estimators are in very close agreement while the power slopes are maintained and coincide with those analytically derived by Valero and Bung (2018). This result highlights that robust techniques allow accurate determination of the turbulence spectrum independently of the employed filtering method.

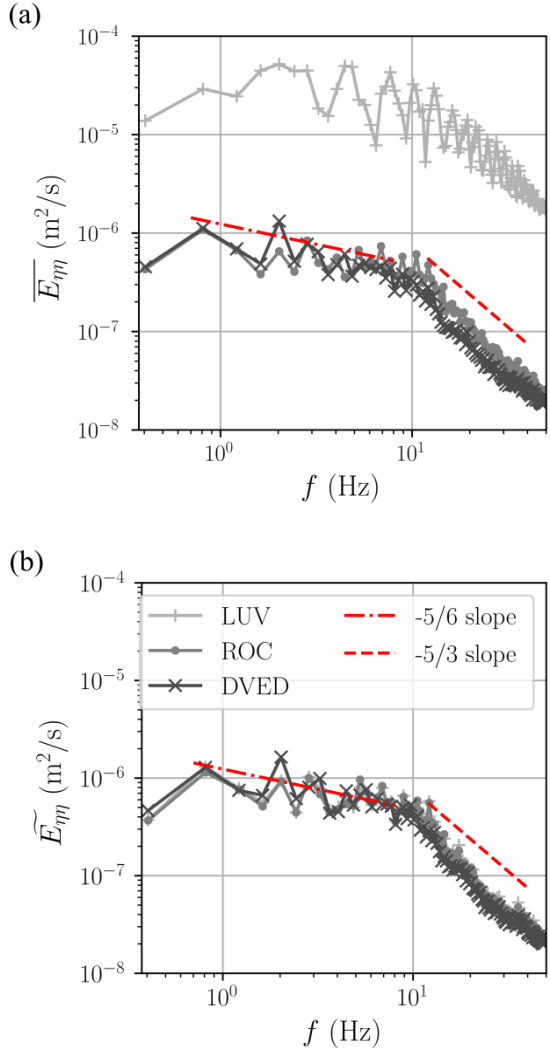

Figure 9. One-dimensional free surface spectrum based on: (a) standard estimators ($\overline{E_{\eta\eta}}$) or (b) robust estimators ($\widetilde{E_{\eta\eta}}$); all averaged over 60 non-overlapping spectra. Data corresponding to one recording of 600 s of ADM 2 for step V and $d_c/h = 1.7$.

## 6 Discussion

A summary of the filtering techniques performance and the effect of robust estimators on most of the studied variables is presented in Tables 2 and 3 and in Fig. 10. The coefficient of determination, defined as the squared value of the Pearson's product moment (Bennett et al. 2013), is used to assess the efficiency of the proposed techniques. This coefficient ranges from 0 to 1, with 0 for no correlation and 1 for perfect correlation. Nonetheless, this efficiency estimator cannot detect bias, which is better observed in Figs. 7 and 8. Raw data has been also included in Fig. 10 for completeness.

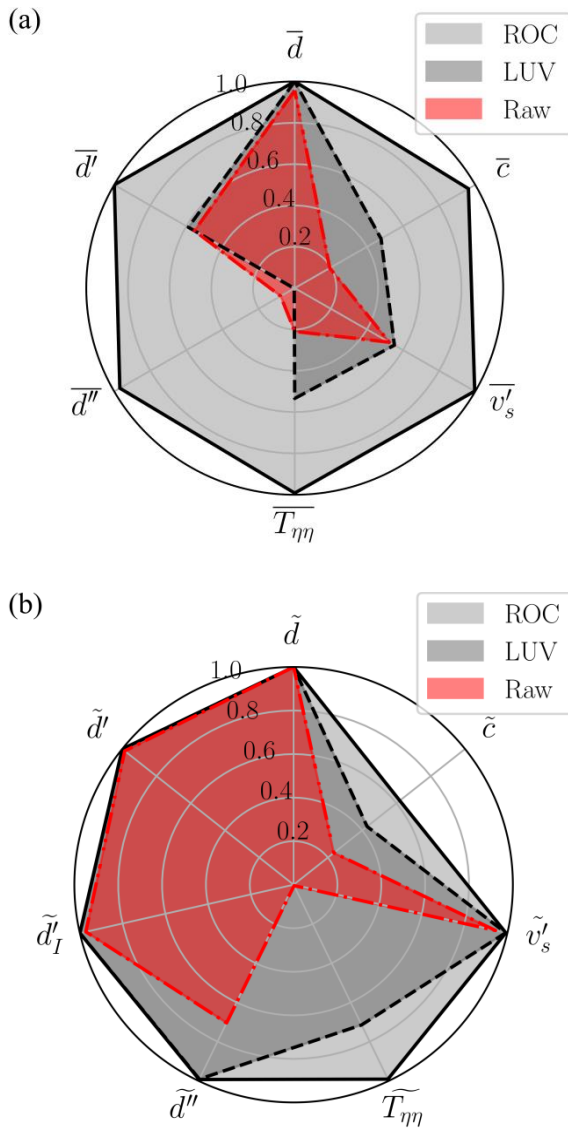

Figure 10. Coefficient of determination for different flow variables using the raw data, LUV filtered data (--) and ROC filtered data (−) against the DVED filtered data. (a) Traditional estimators and (b) alternative robust estimators.

In Fig. 10, the performance is defined against the DVED filtered data, being the most restrictive filtering technique. Nonetheless, the ROC technique achieves similar results with just half the amount of the rejected data (see Fig. 6). Both the raw and the LUV filtered data can yield similarly accurate estimations of expected depths and variance, when used together with robust estimators (MED and MAD). For the skewness determination, the LUV filtering method and robust estimators are the minimum data processing level which should be used. Nevertheless, a similar degree of complexity is involved when using the ROC method and the amount of filtered data does not increase considerably. For the one-dimensional spectrum, the proposed robust non-parametric method allowed detection of the power law scaling whereas power levels converged even with lowest levels of filtering. Robust estimators should be used when possible to reduce the uncertainty of the turbulence predictions, as they are insensitive to the presence of outliers and perform better than the classic estimators for smaller samples size. The only exception observed in this study is the case of the waves' celerity determination.

Table 2. Coefficient of determination (Bennett et al. 2013) for different turbulence quantities. Combination of traditional indicators with raw data, LUV and ROC filtered data. DVED filtered data used as reference.

| Filtering | Variable estimated | | | | | |
|---|---|---|---|---|---|---|
| | $\bar{d}$ | $\overline{d'}$ | $\overline{d''}$ | $\overline{T_{\eta\eta}}$ | $\overline{v_s'}$ | $\bar{c}$ |
| Raw | 0.957 | 0.555 | 0.077 | 0.210 | 0.524 | 0.194 |
| LUV | 1.000 | 0.589 | 0.003 | 0.534 | 0.552 | 0.478 |
| ROC | 1.000 | 1.000 | 0.969 | 0.992 | 0.997 | 0.962 |

Table 3. Coefficient of determination (Bennett et al. 2013) for different turbulence quantities. Combination of robust indicators with raw data, LUV and ROC filtered data. DVED filtered data used as reference.

| Filtering | Variable estimated | | | | | | |
|---|---|---|---|---|---|---|---|
| | $\tilde{d}$ | $\widetilde{d'}$ | $\widetilde{d'_I}$ | $\widetilde{d''}$ | $\widetilde{T_{\eta\eta}}$ | $\widetilde{v_s'}$ | $\tilde{c}$ |
| Raw | 1.000 | 0.985 | 0.972 | 0.705 | 0.005 | 0.944 | 0.235 |
| LUV | 1.000 | 1.000 | 0.999 | 0.991 | 0.715 | 0.992 | 0.429 |
| ROC | 1.000 | 1.000 | 1.000 | 0.993 | 0.991 | 0.993 | 0.619 |

## 7 Conclusions

Classic and robust estimators together with three different filtering techniques have been investigated aiming to shed some light on the accurate determination of free surface turbulent quantities. Measurements were conducted in the non-aerated region of a large stepped spillway model (Fig. 1), which represents one of the most challenging turbulent free surface flows investigated in literature. The performances of the investigated classic and robust estimators and filtering techniques were assessed for the most common turbulence flow variables, ranging from simple fluctuation intensities to the autocorrelation timescales and one-dimensional turbulence spectrum.

The findings show that, when using classic estimators, filtering based on the ROC method is a minimum to provide sound turbulence estimations. The results did not significantly change with the application of the most restrictive filtering method (DVED), although the amount of rejected data was doubled. The use of robust estimators has been explored, thus providing with an alternative tool, insensitive to the presence of outliers. For all studied turbulent variables, at least a resilient estimator has been proposed. Using robust estimators, the turbulence predictions remained accurate even when the filtering technique would be insufficient with classic estimators.

**Funding**

The presented results were obtained in the framework of a project-related personal exchange program funded by the German Academic Exchange Service (DAAD) and Australia-Germany Joint Research Cooperation scheme with financial support of the Federal Ministry of Education and Research (BMBF).

**Notation**

| | |
|---|---|
| ADM | Acoustic Displacement Meter |
| DVED | Depth-Velocity Elliptical Despiking |
| $d$ | Flow depth time series (m) |
| $\bar{d}$ | Mean flow depth (m) |
| $\tilde{d}$ | Median flow depth (m) |
| $d'$ | Flow depth expected deviation (m) |
| $\bar{d'}$ | Flow depth standard deviation (m) |
| $\tilde{d'}$ | Flow depth robust standard deviation ($k\,\mathrm{MAD}(d)$) (m) |
| $\widetilde{d_I'}$ | Flow depth interquartile range (m) |
| $\widetilde{d_n'}$ | Flow depth relative to quartile $n^{\text{th}}$ (m) |
| $\overline{d''}$ | Flow depth expected skewness (-) |
| $\widetilde{d''}$ | Flow depth quartile coefficient of skewness (-) |
| $E_{\eta\eta}$ | One-dimensional flow depth spectra (m$^2$s$^{-1}$) |
| $\overline{E_{\eta\eta}}$ | One-dimensional flow depth spectra, classic determination (m$^2$s$^{-1}$) |
| $\widetilde{E_{\eta\eta}}$ | One-dimensional flow depth spectra, robust determination (m$^2$s$^{-1}$) |
| $f$ | Frequency (s$^{-1}$) |
| $k$ | Relation between the Standard Deviation and the Median of the Absolute Deviation |
| LUV | Lower and Upper Voltage |
| MAD | Median Absolute Deviation |
| MED | Median |
| $N$ | Number of data points (-) |
| PMF | Probability Mass Function |
| $\overline{R_{max}}$ | Maximum cross-correlation, obtained using classic correlation (-) |
| $\widetilde{R_{max}}$ | Maximum cross-correlation, obtained using Spearman's rank correlation (-) |
| $R_{\eta\eta}$ | Autocorrelation function for $\eta$ (-) |
| $\overline{R_{\eta\eta}}$ | Classic autocorrelation function for $\eta$ (-) |
| $\widetilde{R_{\eta\eta}}$ | Autocorrelation function for $\eta$ by means of Spearman's rank correlation (-) |
| ROC | Robust Outlier Cutoff |
| STD | Standard Deviation |
| $T_{\eta\eta}$ | Turbulent timescale for the free surface fluctuation (s) |
| $\overline{T_{\eta\eta}}$ | Turbulent timescale for the free surface fluctuation obtained using the Pearson's product moment correlation (s) |
| $\widetilde{T_{\eta\eta}}$ | Turbulent timescale for the free surface fluctuation obtained using the Spearman's rank correlation (s) |
| $t$ | Time (s) |
| $u$ | Ranked vector (-) |
| $v_s$ | Vertical velocity of the free surface (ms$^{-1}$) |
| $\overline{v_s'}$ | Vertical velocity standard deviation (ms$^{-1}$) |
| $\overline{v_s'}$ | Vertical velocity standard deviation (ms$^{-1}$) |
| $\widetilde{v_s'}$ | Vertical velocity robust standard deviation ($k$ MAD($v_s$)) (ms$^{-1}$) |
| $w$ | Ranked vector (-) |
| $x$ | Auxiliary vector (-) |
| $Z$ | Auxiliary vector (-) |
| $\eta$ | Time series for the deviation from the median flow depth (m) |
| $\lambda_u$ | Universal threshold (-) |